# Competing antiferroelectric and ferroelectric interactions in NaNbO$_3$ : Neutron diffraction and theoretical studies

S.K. Mishra, N. Choudhury, S.L. Chaplot, P.S.R. Krishna and R. Mittal
Solid State Physics Division, Bhabha Atomic Research Centre, Trombay, Mumbai-400085, India.

**Abstract**

Neutron diffraction studies using powder samples have been used to understand the complex sequence of low temperature phase transitions of NaNbO$_3$ in the temperature range from 12 K-350 K. Detailed Rietveld analysis of the diffraction data reveal that the antiferroelectric to ferroelectric phase transition occurs on cooling around 73 K while the reverse ferroelectric to antiferroelectric transition occurs on heating at 245 K. However, the former transformation is not complete till down to 12 K and there is unambiguous evidence for the presence of the ferroelectric *R3c* phase coexisting with an antiferroelectric phase (*Pbcm*) over a wide range of temperatures. The coexisting phases and reported anomalous smearing of the dielectric response akin to dipole glasses and relaxors observed in the same temperature range are consistent with competing ferroelectric and antiferroelectric interactions in NaNbO$_3$. We have carried out theoretical lattice dynamical calculations which reveal that the free energies of the antiferroelectric *Pbcm* and ferroelectric *R3c* phases are nearly identical over a wide range of temperature. The small energy difference between the two phases is of interest as it explains the observed coexistence of these phases over a wide range of temperature. The computed double well depths and energy barriers from paraelectric *Pm$\bar{3}$m* to antiferroelectric *Pbcm* and ferroelectric *R3c* phases in NaNbO$_3$ are also quite similar, although the ferroelectric *R3c* phase has a slightly lower energy.

**I. INTRODUCTION**

Sodium niobate based ceramics exhibit interesting electrical and mechanical properties which find important technological applications[1-5]. Material engineering of alkaline niobates like potassium sodium niobate and lithium sodium niobate with ultralarge piezoresponse comparable to Pb (Zr$_x$Ti$_{1-x}$)O$_3$ (PZT) have evoked considerable interest as the next generation ecofriendly lead free piezoceramics[2,3] required for applications as transducers in cell phones, medical implants, *etc*. Diffuse scattering and relaxor type behavior in NaNbO$_3$ based solid solutions have also been recently reported [6,7].

Pure NaNbO$_3$ has an antiferroelectric (AFE) structure at room temperature and exhibits an unusual complex sequence of temperature and pressure driven structural phase transitions[8-30] which are not clearly understood. The antiferroelectric materials are essentially non-polar but revert to a ferroelectric (FE) polar state when subjected to an external field. Sodium niobate NaNbO$_3$ is a well documented antiferroelectric where electric field switching to the ferroelectric form has been obtained for realizable electric fields. This property makes it particularly important for various material applications including high-density optical data storage.[1] NaNbO$_3$ is also used in enhancing nonlinear optical properties and finds applications in Hologram recording materials.[1,5]

The complex sequence of phase transitions of NaNbO$_3$ is summarized in Fig. 1. Experimental studies of NaNbO$_3$ using a wide variety of techniques including x-ray and neutron diffraction[8-15], Raman and infrared spectroscopy[16-21], XAFS[22], transmission electron microscopy (TEM)[23], electron paramagnetic resonance (EPR)[24], nuclear magnetic resonance (NMR)[25], *etc*. have been reported and the crystal structure[8-15], phonon spectrum[16-21,26], dielectric and piezoelectric behavior[4,6,26-30], temperature and pressure driven phase transitions[8-30] and critical behavior of the order parameter[24,25] studied. First principles calculations of the structural and dynamical properties, epitaxial strain and anomalous LO-TO splittings of NaNbO$_3$ have also been reported[31-36] In spite of the extensive experimental[4-30] and theoretical[31-36] studies, there are numerous controversies surrounding the phase diagram of NaNbO$_3$ with conflicting reports in the literature[29] on the existence of ferroelectric ordering at low temperature. Recent x-ray diffraction and Raman spectroscopic studies suggest that the observation of noncentrosymmetric ferroelectric phases in NaNbO$_3$ is strongly influenced by the particle size[37], but these are again not clearly understood.

As material properties are strongly influenced by the associated crystal structure, accurate characterization of the phase diagram of NaNbO$_3$ is essential for design of new sodium niobate based ceramics for applications. In the present work, we have employed systematic neutron diffraction measurements as a function of temperature (T=12 K to 350 K) to study the low temperature structures and understand their phase transitions. Neutron diffraction offers certain unique advantages over x-rays especially in the accurate determination of oxygen positions and provides information about subtle changes in structure (associated with the oxygen atoms) accompanying these phase transitions, which are crucial for resolving the existing controversies. Our neutron diffraction studies reveal that the ferroelectric and antiferroelectric phases coexist over a wide temperature range. The coexistence of both phases and reported anomalous smearing of dielectric response over the same temperature range in NaNbO$_3$[6,18] are consistent with competing ferroelectric and antiferroelectric interactions.[38,39]

The role of soft phonon modes in driving the complex sequence of phase transition is of additional interest. Using first principles total energy calculations, Zhong and Vanderbilt[32] show that the instabilities associated with zone-center polar ferroelectric mode and the zone boundary non-polar antiferrodistortive modes lead to a very rich phase diagram. In the present work, we have undertaken theoretical calculations using an optimized model aimed at understanding (i) the double well depths and nature of the lattice instabilities from the ideal paraelectric to the antiferroelectric and ferroelectric structures, and (ii) the free energies and relative stability of the ferroelectric and antiferroelectric phases. We obtain a very small energy difference between the ferroelectric and antiferroelectric phases which is of interest as it could be overridden by a realizable electric field, which is important for applications.

## II. TECHNIQUES

### A. Experimental

The powder neutron diffraction data were recorded in the $2\theta$ range 7-138 degree at a step of width $0.05^o$ using neutrons of wavelength 1.249 Å on a medium resolution powder diffractometer in the Dhruva Reactor at Bhabha Atomic Research Centre. An 8 *mm* diameter thin-walled vanadium can of length 55 *mm* is used to hold the powder sample and measurements were carried out in a closed cycle refrigerator in heating and cooling cycle (300 K- 12 K – 350 K – 12 K) at various temperatures. The accuracy of the measured temperature during data collection was within ±0.1 K at the cold head end. The temperature gradient across the length of the sample was measured to be less than 1K at temperature above 50 K but was as much as 5 K at the temperature 12 K. The structural refinements were performed using the Rietveld refinement program FULLPROF.[40] In all the refinements, the background was defined by a sixth order polynomial in $2\theta$. A pseudo-Voigt function was chosen to define the profile shape for the neutron diffraction peaks. Except for the occupancies of the atoms, all other parameters i.e., scale factor, zero correction, background, half-width parameters along with mixing parameters, lattice parameters, positional coordinates, and thermal parameters were refined.

### B. Theoretical

Theoretical calculations of the crystal structure, total energies and dynamics require information about the interatomic forces which can be obtained either by using a quantum-mechanical *ab-initio* formulation or using semiempirical interatomic potentials. The perovskite structure (Fig. 2), consists of a three dimensional network of corner shared $NbO_6$ octahedra with Na atoms occupying the interstices. The antiferroelectric distortions involve octahedral rotations and tilts (Fig. 2) and there are 40 atoms/unit cell in the antiferroelectric *Pbcm* structure and 30 atoms/unit cell in the ferroelectric *R3c* structure. Due to the structural complexity, we have undertaken studies using interatomic potentials involving long ranged Coulombic and short-ranged Born-Mayer type terms, given by

$$V(r) = \frac{e^2}{4\pi\varepsilon_o} \frac{Z(k)Z(k')}{r} + a\, exp\left\{\frac{-br}{R(k)+R(k')}\right\} - \frac{C}{r^6}$$

Here *r* is the separation between the atoms of type *k* and *k'*. $R(k)$ and $Z(k)$ are the effective radius and charge of the $k^{th}$ atom type. We have treated $a$=1822 eV and $b$=12.364 as constants as used earlier in the lattice dynamical calculations of several complex solids.[41] The calculations are carried out using the code DISPR[42]. The crystal structure at any pressure (at zero temperature) is obtained by minimization of the enthalpy. Our experimental crystal structure and reported Raman[16-20] and infrared[26] data have been used to refine the interatomic potentials. Using these potentials, we have studied the double wells and computed the zero pressure free energies including the complete vibrational contributions[43], to understand the relative stability of the ferroelectric *R3c* and antiferroelectric *Pbcm* phases.

## III. RESULTS

### A. Evolution of the ferroelectric to antiferroelectric phase transition

Figure 3 depicts the evolution of powder neutron diffraction patterns of $NaNbO_3$ with temperature in the range of 12–350 K for cooling and heating cycles. $NaNbO_3$ has an antiferroelectric phase with space group *Pbcm* at room temperature (300 K). The antiferroelectric phase consists of two or more sublattice polarizations of antiparallel nature, which in turn give rise to superlattice reflections in the diffraction patterns. We have observed such reflections in the antiferroelectric phase and the strongest one of them appears at $2\theta \approx 29.5$ degree. These reflections assume the indices *h k l/p* (*h,k* and *l* are integers, *p* gives the multiple along [001] direction) with respect to the equivalent elementary perovskite cell. These characteristic antiferroelectric reflections are marked with arrows. All the peaks in the powder neutron diffraction patterns at 300 K, could be indexed using *Pbcm* space group.

It is evident from the Fig. 3, that there is a drastic change in the profile near $2\theta \approx 32$, 42 and 55 degrees in the diffraction pattern below 100 K during cooling and above 225 K in heating cycles, respectively. The intensity of characteristic antiferroelectric peaks ($2\theta \approx 29.5$ and 44.4 degrees) decreases gradually on lowering the temperature while the peaks at $2\theta \approx 32$, 44 and 55 degrees show splitting.

## B. Selection of correct space group at low temperature

From the knowledge of the Miller indices of the superlattice reflections and lattice type of the structure, as revealed by the splitting of main Bragg peaks, one can drastically restrict the number of the plausible space groups to be considered in the Rietveld analysis. Rhombohedral distortion, would lead to the splitting of the *hhh* type pseudocubic reflections into two while the *h00* type reflections would remain singlet. Thus, the splitting in the peaks at 2θ ≈ 32 degree (222 reflection) is unambiguous signature of the presence of rhombohedral symmetry in the ferroelectric phase (N) at low temperature.

The integrated intensity of the characteristic antiferroelectric peak (at 2θ ≈29.5 degrees) as a function of temperature is shown in figure 4 (a), for the cooling and heating cycles. Figures 3 and 4 (a) suggest that both antiferroelectric and ferroelectric phase are present at 12 K. We therefore, refined the structure at 12 K using the space groups *Pbcm* (Sp. Gr. No. 57) and *R3c* (Sp. Gr. No. 161) by the Rietveld technique. For the *Pbcm* space group with orthorhombic symmetry we used $A_o=a_p+c_p$, $B_o=a_p-c_p$ and $C_o=4b_p$ axis, where $a_p$, $b_p$ and $c_p$ refer to the elementary perovskite cell. The asymmetric unit of the structure consists of two Na atoms, Na1 at the *4c* site (1/4+$u$, ¾, 0) and Na2 at the *4d* site (1/4+$u$, ¾ +$v$, 1/4); one Nb atom at the *8e* site (1/4+$u$, 1/4+$v$, 1/8+$w$) and four O atoms, O1 at the *4c* site (1/4+ $u$, 1/4, 0), O2 at the *4d* site (1/4+$u$, 1/4 +$v$, 1/4), O3 at the *8e* site (1/2+$u$, 0+$v$, 1/8+$w$) and O4 at the *8e* site (0+$u$, 1/2+$v$, 1/8+$w$), respectively. In the space group *R3c*, for the description of the crystal structure we used the following hexagonal axes: $a_h=b_p-c_p$, $b_h=c_p-a_p$ and $c_h=2a_p+2b_p+2c_p$. The hexagonal unit cell contains six formula units, while the true unit cell in the primitive rhombohedral contains two formula units. The asymmetric unit of the structure consists of Na atom at the *6a* site at (0, 0, ¼+$w$), Nb atom at the *6a* site at (0,0,0+$w$) and O atom at the *18b* site at (1/6+$u$, 1/3+$v$, 1/12). The refinements converged smoothly. In Figure 4(b), we display the observed, calculated and difference profiles for *Pbcm*+*R3c* at 12 K. The fit between the observed and calculated profiles is quite satisfactory and include the weak antiferroelectric superlattice reflection as can be seen from the inset in Fig. 4(b). The refined structural parameters are given in Table I.

## C. Variation of phase fraction and lattice parameters with temperature

The results of the preceding section show that in NaNbO$_3$, the antiferroelectric *Pbcm* and ferroelectric *R3c* phases coexist over a wide range of temperatures. The phase fraction as a function of temperature obtained by Rietveld refinement is plotted in Fig. 5 (a) for the cooling and heating cycles, respectively. After cooling, the sample was retained at 12 K for 6 hours. On increasing the temperature from 12 K, the phase fraction of the antiferroelectric phase (~22 %) is nearly constant up to 200 K and then shows an abrupt increase around 245 K. Rietveld refinement of the neutron data (Fig. 5a) clearly reveals that for T= 275 K, the majority phase is antiferroelectric (>99%) while the minority phase is ferroelectric (< 1%). The variations of the lattice parameters with temperature for the both phases are shown in Fig. 5 (b) and (c). For the sake of easy comparison with corresponding cell parameters of orthorhombic and rhombohedral phases, we have plotted the cell parameters of these phases in terms of equivalent elementary perovskite cell parameters. Figure 5(b) reveals that with increasing temperature, the cell parameters $a_p$ and $c_p$ of the antiferroelectric phase (*Pbcm*) increases monotonically, whereas $b_p$ shows decreasing trends. The lattice parameter of NaNbO$_3$ in the rhombohedral (*R3c*) phase monotonically increases with temperature and also show a slope change around 245 K. This provides the evidence for the phase transition from the ferroelectric to the antiferroelectric phase in heating cycle (Fig 5c). In the cooling cycle, the phase reversal occurs at around 73 K (see Fig. 5).

## D. Theoretical calculations of the double wells and lattice instabilities

Using the theoretical model described in Section II. B, we have calculated the energy barriers between the paraelectric, ferroelectric *R3c* and antiferroelectric *Pbcm* phases. The calculated zero pressure structural parameters are in good agreement with the experimental diffraction data The paraelectric cubic $Pm\bar{3}m$ structure has a higher energy than the ferroelectric *R3c* and antiferroelectric *Pbcm* phases, and we have calculated the double wells corresponding to the ferroelectric and antiferroelectric distortions (Fig. 6 (a)). Our calculations reveal that both the ferroelectric and antiferroelectric distortions in NaNbO$_3$ yield similar lowering of the energy as compared to the higher energy paraelectric phase, although the ferroelectric phase has a slightly lower energy.

Our experimental and calculated results are overall consistent with first-principles calculations[34, 35] of NaNbO$_3$ which suggest that the ground state is characterized by ferroelectric atomic displacements and frozen tilting of oxygen octahedra corresponding to the *R* point instabilities of the Brillouin zone. Although the ferroelectric *R3c* phase has a slightly lower internal energy (Fig. 6), the slightly higher vibrational entropy of the antiferroelectric *Pbcm* phase causes the free energy crossover at $T_c$ ~ 50 K. The free energy differences of the antiferroelectric *Pbcm* and ferroelectric *R3c* structure (see inset of Fig. 6(b)) are however within thermal fluctuations in the 0-400 K temperature range which explain the coexisting ferroelectric and antiferroelectric structures over wide range of temperature in NaNbO$_3$ observed in our experiments.

The ferroelectric *R3c* phase involves polar displacements of the Na, Nb and O atoms along the pseudocubic [111] direction. Using displacements of atoms from ideal positions and the reported averaged values of Born effective charge tensors reported for NaNbO$_3$ from first-principles calculations[35], we have estimated the spontaneous polarization of the ferroelectric *R3c* structure. The calculated spontaneous polarization of NaNbO$_3$ for the observed *R3c* structure given in Table I is found to be 0.59 C/m$^2$ which is comparable to that observed for a conventional ferroelectric like tetragonal PbTiO$_3$ (0.57 C/m$^2$) and significantly smaller than LiNbO$_3$ (0.71 C/m$^2$).[1]

## V. DISCUSSION

The powder pattern in the ferroelectric phase has less number of reflections compared to the antiferroelectric phase, as the result of the change in the multiplicity of unit cell. The room temperature antiferroelectric phase (P phase) has alternating pairs of $\bar{a}\bar{a}\bar{b}^+$ and $\bar{a}\bar{a}\bar{b}^-$ layers. These are associated with R (½,½,½) and Δ (0,0,¼) points instability of cubic Brillouin zone. In the transition from antiferroelectric (P phase, cell multiplicity 2×2×4 in terms of pseudocubic perovskite) to the rhombohedral ferroelectric phase (N phase) the tilting scheme changes to $\bar{a}\bar{a}\bar{a}$ and unit cell multiplicity becomes 2×2×2. Similarly, the ferroelectric phase has less number of Raman lines than antiferroelectric phase. Since, at lowest temperature, ferroelectric phase is the majority phase, the intensities of Raman active modes are mostly accounted by this phase and as a result, various workers [4(b), 16 18, 20] assigned Raman lines by this single phase only. Shen *et al*[20] have shown that ferroelectric to antiferroelectric phase transition occurs around 210 K on heating. They have also found that both the ferroelectric and antiferroelectric phases coexist down to 43 K. However, they found that the Raman lines could be explained by purely ferroelectric phase at 10 K. Recently, Yuzyuk *et al*[4(b)] using synchrotron study, have also showed that ferroelectric to antiferroelectric phase transition occurs at 250 K on heating. All these results are in good agreement with our findings.

During cooling cycle we find that the antiferroelectric to ferroelectric phase transition occurs around 73 K, and our results are consistent with the temperature dependent Raman spectroscopic studies of Lima *et al*.,[18] who obtain an AFE to FE phase transition around 95 K. Detailed analysis of our powder neutron diffraction data reveal that the new unidentified phase between the ferroelectric and antiferroelectric phases (Fig. 1) obtained from temperature dependent studies of Lima *et al*.,[18] is actually not a new phase but a coexistence of the ferroelectric and antiferroelectric phases which can explain the anomalous behavior in the observed Raman spectra[18] in this temperature range.

Competing ferroelectric and antiferroelectric interactions in the Sr$_{1-x}$Ca$_x$TiO$_3$ (SCT) system[38,39] were shown to lead to smearing of the dielectric response similar to dipole glasses and relaxor ferroelectrics. Diffuseness and smearing of dielectric response have also been reported in NaNbO$_3$ over the temperature range [80 K-250 K] by Lanfredi and coworkers.[6,18] This reported anomalous dielectric response occurs over the same temperature range over which we obtain coexisting ferroelectric and antiferroelectric phases. As explained earlier for SCT,[38, 39] competing ferroelectric and antiferroelectric interactions can explain the diffuse smearing in the reported observed anomalous dielectric response of NaNbO$_3$.

To understand the origin of the coexisting phases and competing interactions, we undertook theoretical lattice dynamical calculations described in detail in section II.B. Our studies suggest that the free energy difference between the ferroelectric and antiferroelectric phases is small and within thermal fluctuations, explaining their coexistence over a wide temperature range in the experiments. This result is significant, as for application purposes, only those antiferroelectrics which are close in free energy to alternative ferroelectric forms are of interest as their energy difference can then be over ridden by a realizable electric field. Sodium niobate is a well documented antiferroelectric where electric field switching to the ferroelectric form has been obtained for realizable electric fields[1,44] which make it useful for applications.

## V. Conclusions

Powder neutron diffraction studies of NaNbO$_3$ have been carried out in the temperature range 12 K-350 K to accurately characterize the low temperature crystal structure and understand the phase transitions. The results are able to resolve and explain contrary reports on the characterization of phases from previous Raman and X-ray diffraction experiments below 300 K. We obtain an antiferroelectric to ferroelectric phase transition on cooling around 73 K and the reverse transition at 245 K, with the FE and AFE phases coexisting over a wide temperature range from 12- 280 K. Theoretical calculations using interatomic potentials suggest that the free energies of the ferroelectric and antiferroelectric phases over this temperature range are quite close. The small energy difference between the two phases is of interest as it would make it possible to easily switch from the antiferroelectric to ferroelectric state using realizable electric fields, which in turn, would determine the potential use of this material for applications. The calculations provide estimates of the spontaneous polarization and help understand the coexistence of the phases. Our studies suggest that competing ferroelectric and antiferroelectric interactions can explain the reported observed[6] anomalous smeared dielectric response akin to dipole glasses and relaxors observed in NaNbO$_3$ over the 12-280 K temperature range.

*Acknowledgements:* SKM acknowledges the help of Mr. A.B. Shinde during the course of neutron diffraction experiments.

**References**
1. M. E. Lines and A. M. Glass "Principles and Application of Ferroelectrics and Related Materials" (oxford: Clarendon, 1977); Xu. Yuhuan, Ferroelectric Materials and Their Applications (Nort-Holland Elsevier Science, 1991); L. G. Tejuca and J. L. G. Fierro " Properties and Applications of Perovskite-Type Oxides" (New York: Dekker, 1993).
2. Y. Saito, H. Takao, T. Tani, T. Nonoyama, K. Takatori, T. Homma, T. Nagaya, M. Nakamura, Nature **423**, 84 (2004).
3. E. Cross, Nature **432**, 24 (2004).
4. (a) Yu. I. Yuzyuk, P. Simon, E. Gagarina, L. Hennet, D. Thiaudiere, V. I. Torgashev, S. I. Raevskya, I. P. Raevskii, L. A. Reznitchenko and J. L. Sauvajol, J. Phys.: Condens. Matter **17**, 4977 (2005);(b) Yu.I. Yuyuk, E. Gagarina, P. Simon, L A Reznitchenko, L Hennet and D. Thiaudiere, Phys. Rev B **69**, 144105 (2004).
5. E Valdez, C B de Araujo, A. A. Lipovskii Appl. Phys. Lett., **89**, 31901 (2006); E Hollenstein, M. Davis, D. Damjanovic and Nava Setter Appl. Phys. Lett., **87**, 182905 (2006); E. L. Falcão-Filho, C. A. C. Bosco, G. S. Maciel, L. H. Acioli, and Cid B. de Araújo, A. A. Lipovskii, D. K. Tagantsev, Phys. Rev. B **69**, 134204 (2004); G. S. Maciel, N. Rakov, Cid B. de Araujo, A. A. Lipovskii and D. K. Tagantsev, Applied Physics Lett. **79**, 5 584 (2001).
6. S. Lanfredi, M.H. Lente and J.A. Eiras, Appl. Phys. Lett., **80**, 2731 (2002); M. H. Lente J. de Los S. Guerra, J. A. Eiras and S. Lanfredi, Solid State Commun. **131**, 279 (2004).
7. I P Raevski, S I Raevskaya, S A Prosandeev, V A Shuvaeva, A M Glazer and M S Prosandeeva, J. Phys.: Condens. Matter 16, L221-L226( 2004) ;I.P. Raevski, L.A. Reznitchenko, M A Malitskaya, L A Shilkina, S.O. Lisitsina, S.I. Raevskaya, E. M. Kuznetsova, Ferroelectrics **299**, 95 (2004); A. I. Burkhanov, P.V. Bondarenko, S I Raevskaya, A.V. Shil'nikov and I P Raevski, Physics of the Solid State, **48**, 1114 (2006); I. P. Raevski, S. A. Prosandeev, Lubomír Jastrabík, Integrated Ferroelectrics, **47**, 277-283 1058 (2002).
8. C. N. W. Darlington and K. S. Knight, Physica B **266**, 368 (1999); C. N. W. Darlington and K. S. Knight, Acta. Cryst., B **55**, 24 (1999).
9. C. N. W. Darlington and H. D. Megaw, Acta. Cryst., B **29,** 2171 (1973).
10. A. M. Glazer and H. D. Megaw, Acta. Cryst., A **29**, 489 (1973).
11. M. Ahtee, A M. Glazer, and H.D Megaw, Phil. Mag. **26** 995 (1972).
12. A. M. Glazer, H.D. Megaw, Phil. Mag. **25** 1119 (1972).
13. A.C. Sakowski-Cowely, K. Lukazewicz and H. D. Megaw, Acta Cryst., B **25**, 851 (1969); A.C. Sakowski-Cowely Ph. D. Thesis, University of Cambridge (1967).
14. I. Lefkowitzk, K. Lukazewicz and H. D. Megaw, Acta. Cryst., **20**, 670 (1966).
15. F. Denoyer, M. Lambert and R. Comes, Solid State Commun., **18**, 441 (1976) and reference therein.
16. S. J. Lin, D. P. Chiang, Y. F. Chen, C. H. Peng, H. T. Liu, J. K. Mei, W. S. Tse, T.-R. Tsai, H.-P. Chiang, Journal of Raman Spectroscopy **37**, 1442 (2006).
17. E. Bouizane, M. D. Fontana and M. Ayadi, J. Phys.; Condens. Matter, **15**, 1387 (2003).
18. R. J. C. Lima, P. T. C. Freire, J. M. Saski, A. P. Ayala, F. E. A. Melo, J. Mendes Filho, K. C. Serra, S. Lanfredi, M. H. Lente and J. A. Eiras J. Raman Spectrosc., **33**, 669 (2002).
19. Shen, Z X, XB W ang, S H Tang, M H Kuok and R Malekfar, J. Raman Spectrosc. **31**, 439 (2000).
20. Z. X. Shen, X. B. Wang, M. H. Kuok, and S. H. Tang, J. Raman Spectrosc., **29**, 379 (1998).
21. X. B Wang, Z. X. Shen, Z. P. Hu, L. Qin, S. H. Tang, M. H. Kuok, J. Molecular Structure, **385**, 1 (1996).
22. V. A. Shuvaeva, Y. Azuma, K. Yagi, K. Sakaue and H. Terauchi, *J. Synchrotron Rad.* **8**, 833, (2001).
23. J. Chen and D. Feng, Phys. Stat. Sol. A. **109,** 171 (1988), J. Chen and D. Feng, *ibid*, **109,** 427 (1988).
24. A. Avogadro, G Bonera, F. Borsa and A. Rigamont, Phys. Rev. B **9**, 3905 (1974).
25. G. Adriano, S. Aldrovandi and A. Rigamonti, Phys Rev B **25,** 007044 (1982); S. E. Ashbrook, L. Le Polles, R. Gautier, C.J. Pickard, R. I. Walton, Physical Chemistry Chemical. Phys., **8**, 3423 (2006).
26. F. Gervais, J. L. Servoin, J. F. Baumard and F. Denoyer, Solid State Commun., **41,** 345 (1982).
27. L. A. Reznitchenko, A. V. Turik, E. M. Kuznetsova and V. P. Sakhnenko, J. Phys.: Condens. Matter, **13**, 3875 (2001).
28. A. Reisman, F. Holtzberg and E. Banks, J Amer. Chem. Soc. **80**, 37 (1958).
29. G. Shirane, R. E. Newman and R. Pepinsky, Phys. Rev. **96,** 581 (1954) and reference therein.
30. L. E. Cross and B. J. Nicolson, Phil. Mag., **46**, 453 (1955).
31. S. Prosandeev, Physics of the Solid State **47**, 11, 2130 (2005).
32. W. Zhong and D. Vanderbilt, Phys. Rev. Lett. **74**, 2587 (1995).
33. O. Diéguez, K. M. Rabe, and D. Vanderbilt, Phys Rev. B **72,** 144101 (2005).
34. D. Vanderbilt and W. Zhong, Ferroelectrics **181**, 206-207 (1998).
35. W.Zhong, R. D. King-Smith and D. Vanderbilt, Phys. Rev. Lett. **72**, 3618 (1994).
36. R. D. King-Smith and D. Vanderbilt, Phys. Rev. B **49**, 5828 (1994).
37. Y. Shiratori, A. Magrez, J. Dornseiffer, F. H. Haegel, C. Pithan, R. Waser, Journal of Physical Chemistry B. **43**, 20122 (2005).
38. R. Ranjan, D. Pandey and N.P. Lalla, Phys Rev Lett. **84,** 3726 (2000).
39. S K Mishra Ph. D. Thesis, Banaras Hindu University (2004).

40. J. Rodriguez-Carvajal, FULLPROF, A Rietveld and pattern Matching Analysis Programm Laboratoire Leon Brillouin (CEA-CRNS) France.
41. S.L. Chaplot, L. Pintschovius, N. Choudhury and R. Mittal, Phys. Rev. B **73**, 094308 (2006); S.L. Chaplot, N. Choudhury, S. Ghose, M.N. Rao, R. Mittal and P. Goel, European Journal of Mineralogy **14**, 291 (2002); R. Mittal, S.L. Chaplot and N. Choudhury, Progress in Materials Science **51**, 211 (2006).
42. S.L. Chaplot (unpublished).
43. S.L. Chaplot, Phys. Rev. B**36**, 8471 (1987); N. Choudhury, S. L. Chaplot, and K. R. Rao, Phys. Rev. B **33**, 8607 (1986); N. Choudhury and S.L. Chaplot, Solid State Comm. **114**, 127 (2000).
44. A..V. Ulinzheev, A.V. Leiderman, V.G. Smotrakov, V. Yu. Topolov and O.E. Fesenko, Phys. Solid State **39**, 972, (1997); R. H. Dungon and R. D. Golding J. Am. Ceram. Soc. **47** (1964).
45. R.T. Downs and M.H. Wallace, Am. Mineral. **88**, 247 (2003).

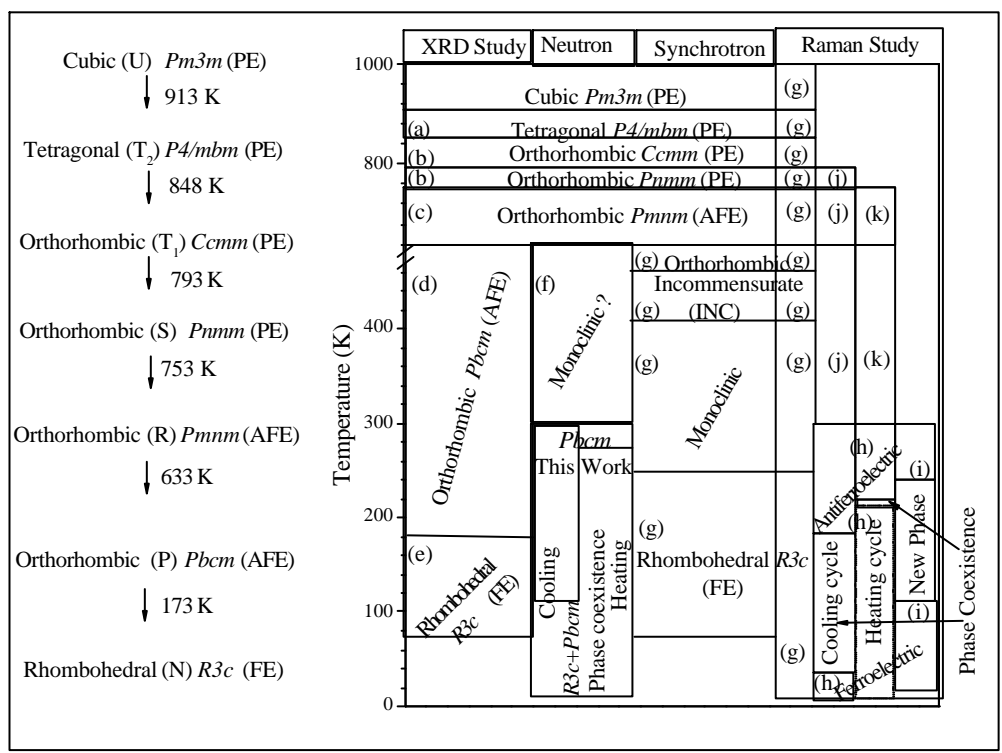

**Fig. 1** Reported phase transitions of NaNbO$_3$ studied using X-ray diffraction (XRD) ([a]Ref. 12, [b]Ref. 11, [c] and [d] Ref. 13 and [e]Ref. 9); neutron diffraction ([f]Ref. 8) and Raman scattering studies ([g] Ref. 4 (a), [h] Ref. 20, [i] Ref. 18, [j] Ref. 21 and [k]Ref. 17). Yuzyuk *et al* ([g] Ref. 4 (a)) have used both synchrotron X-ray diffraction and Raman scattering techniques to study the high temperature phase transitions. The phase transition temperatures obtained using XRD studies[9-13] are indicated in the left. Notations U, T$_2$ *etc.* are from Ref. [8]. Space groups are indicated in italics. FE, AFE and PE correspond to ferroelectric, antiferroelectric and paraelectric phases, respectively.

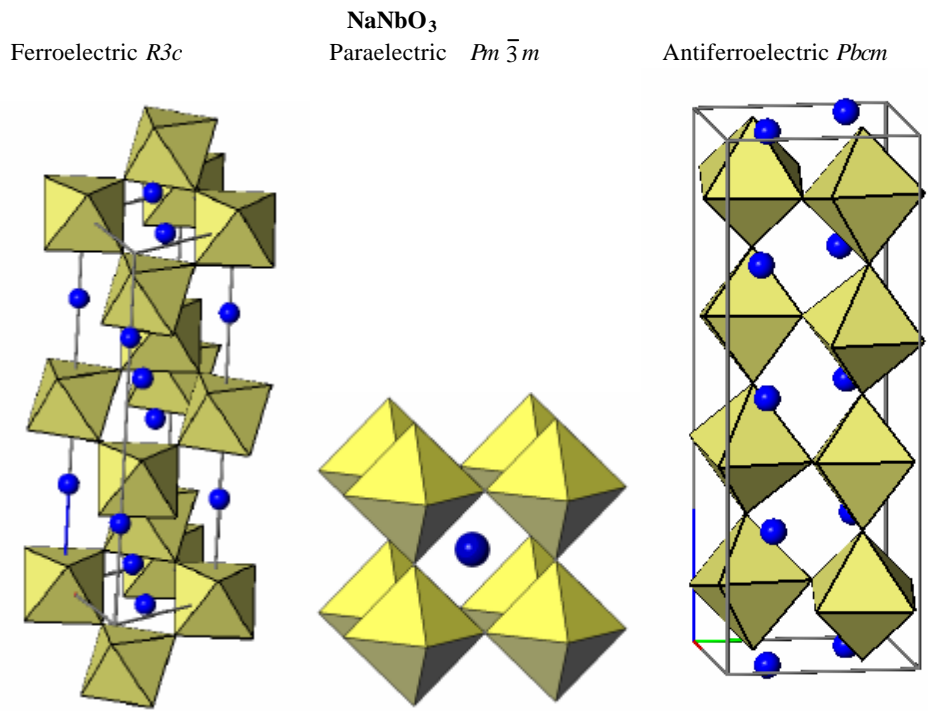

**Fig. 2** (Color online). Polyhedral representation of the ferroelectric *R3c*, paraelectric *Pm $\bar{3}$ m* and antiferroelectric *Pbcm* phases of NaNbO$_3$ displayed using the software xtaldraw [45]. The Na atoms are depicted as spheres.

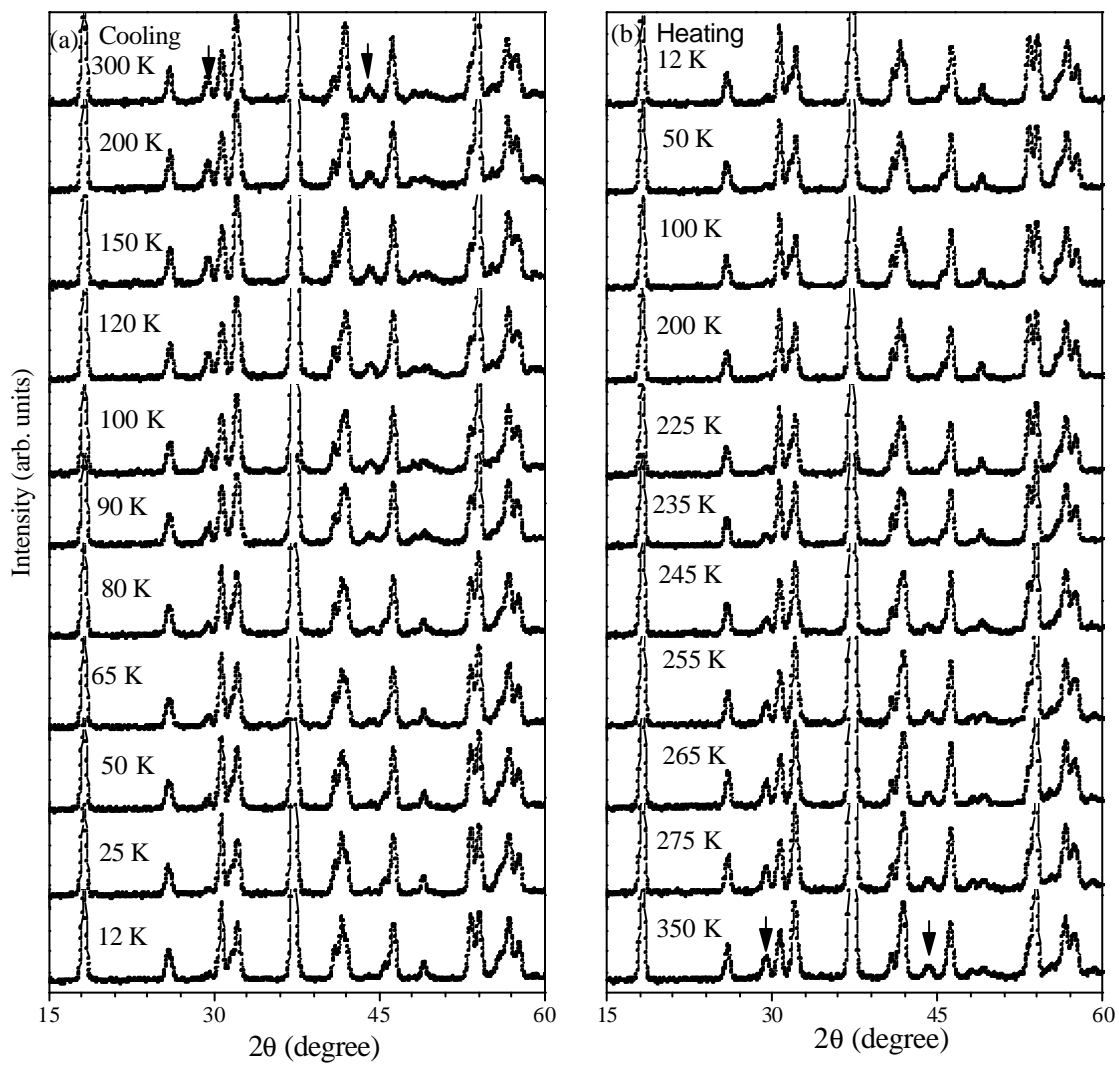

**Fig. 3** Evolution of neutron diffraction patterns with temperature for (a) cooling cycle, and (b) heating cycle. The characteristic antiferroelectric peaks are marked with arrows.

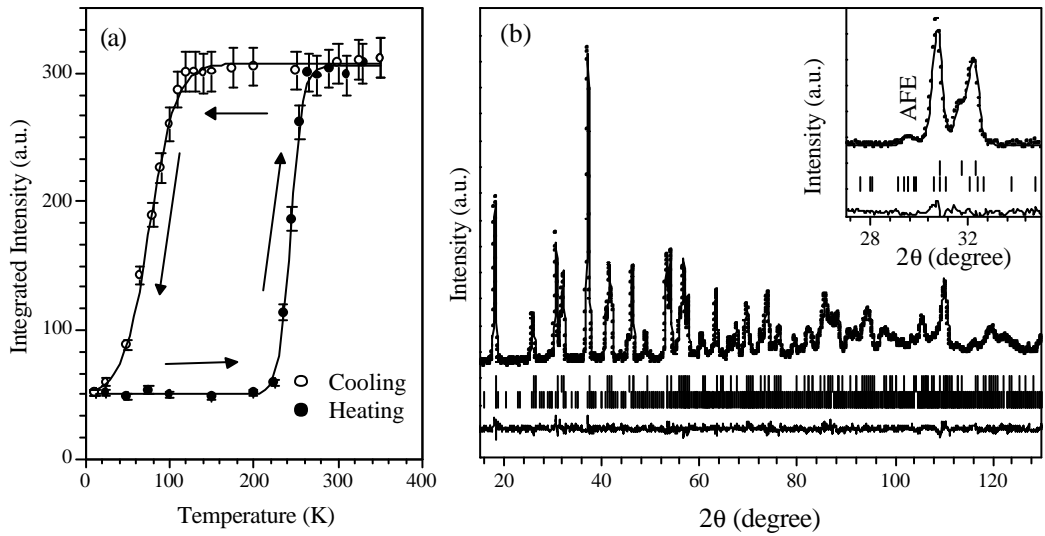

**Fig. 4** (a) Variation of the integrated intensity of antiferroelectric peak appearing around $2\theta=29.5°$ for heating (solid circle) and cooling (open circle) cycles. (b) Observed (dots), calculated (continuous line), and difference (bottom line) profiles obtained after the Rietveld refinement of $NaNbO_3$ in the $2\theta$ range 15-130 degree using two phase (orthorhombic, *Pbcm* and rhombohedral *R3c*) model at 12 K. The inset shows the enlarged patterns for the selected $2\theta$ range to highlight the fit for the weak antiferroelectric peaks (marked with AFE). Upper and lower vertical tick marks above the difference profiles show the peak positions of *R3c* and *Pbcm* phases, respectively.

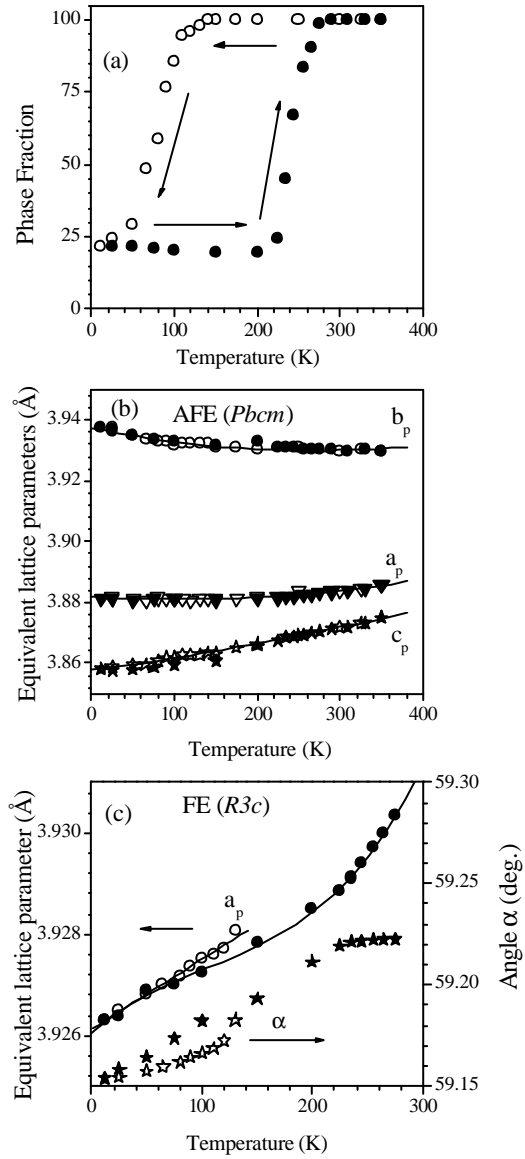

**Fig. 5 (a)** Variation of the phase fraction of the antiferroelectric phase with temperature. Variations of lattice parameters of NaNbO$_3$ in the orthorhombic AFE (b) and rhombohedral FE (c) phases with temperature during heating (solid symbol) and cooling (open symbol) cycles.

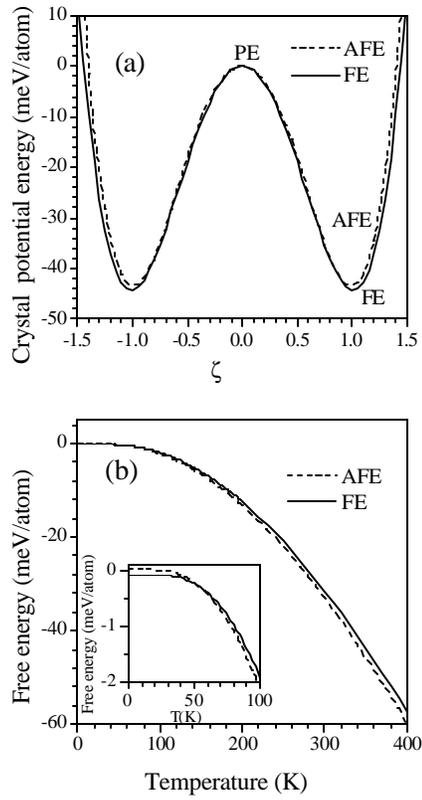

**Fig. 6** (a) Energy barriers from the paraelectric $Pm\bar{3}m$ to antiferroelectric (space group *Pbcm*) phase and ferroelectric (space group *R3c*) phase. ζ corresponds to the symmetry lowering AFE and FE distortions of the ideal cubic paraelectric structure for the dashed and full lines, respectively. (b) Calculated free energy (including vibrational contributions) for the antiferroelectric and ferroelectric phases of $NaNbO_3$.

**Table I.** Refined structural parameters of NaNbO$_3$ at 12 K using the two phase model. The estimated spontaneous polarization of the ferroelectric *R3c* phase is P= 0.59 C/m$^2$ (along the cubic perovskite [111] direction).

---

**Phase 1 Ferroelectric Phase : Space Group *R3c***

Cell parameters: $a_h$=5.48114(3) (Å); $c_h$=13.68518(9) (Å), Vol=356.060(36) (Å)$^3$
Phase Fraction (%): 78.00; $R_B$=2.51, $R_f$=1.25

---

| Atoms | Positional Coordinates | | | Thermal Parameters |
|---|---|---|---|---|
| | X | Y | Z | B (Å)$^2$ |
| Na | 0.000 | 0.000 | 0.2723(4) | 0.2172(12) |
| Nb | 0.000 | 0.000 | 0.0164(2) | 0.3549(17) |
| O | 0.0999(5) | 0.3367(7) | 0.08333 | 0.7680 (21) |

---

**Phase 2 Antiferroelectric Phase : Space Group *Pbcm***

Cell parameters: $A_o$=5.50120(13) (Å); $B_o$=5.56496(13) (Å); $C_o$=15.39720(9) (Å), Vol=471.369(21) (Å)$^3$
Phase Fraction (%): 22.00; $R_B$=3.38, $R_f$=2.00

---

| Atoms | Positional Coordinates | | | Thermal Parameters |
|---|---|---|---|---|
| | X | Y | Z | B (Å)$^2$ |
| Na | 0.247(6) | 0.750 | 0.0000 | 1.3(6) |
| Na | 0.227(3) | 0.789(6) | 0.2500 | 0.2(2) |
| Nb | 0.2416(3) | 0.2819(2) | 0.1314(1) | 0.1(1) |
| O1 | 0.329(2) | 0.250 | 0.0000 | 0.2(1) |
| O2 | 0.208(5) | 0.278(3) | 0.2500 | 0.2(2) |
| O3 | 0.530(3) | 0.0402(2) | 0.1378(2) | 0.9(2) |
| O4 | 0.9751(3) | 0.489(2) | 0.1065(1) | 0.5(2) |

---

Over all fitting parameter $R_p$=4.47, $R_{wp}$=5.83, $R_{exp}$=3.69, $\chi^2$=2.65